\documentclass[twocolumn,aps,floats,nofootinbib,prd,amsmath,amssymb,secnumarabic]{revtex4}
\usepackage{graphicx, natbib, color}

\newcommand{\phie}{\phi_{\rm end}}
\newcommand{\deep}{\partial}
\newcommand{\ns}{n_{\rm s}}

\newcommand{\Ri}{R_\mathrm{i}}
\newcommand{\Rit}{\hat{R}_\mathrm{i}}
\newcommand{\ms}{m_s}
\newcommand{\gs}{g_s}
\newcommand{\eg}{{\sl e.g.}}
\newcommand{\Ps}{P_{\rm s}}
\newcommand{\Pt}{P_{\rm t}}
\newcommand{\MP}{M_{\rm pl}}

\begin{document}
\title{Brane Inflation and the Overshoot Problem}
\author{Simeon Bird$^1$\footnote{spb41@cam.ac.uk}, Hiranya V. Peiris$^1$, and Daniel Baumann$^2$} 
\affiliation{$^1$Institute of Astronomy, University of Cambridge, Cambridge CB3 0HA, U.K.\\
$^2$Department of Physics, Harvard University, Cambridge, Massachusetts 02138, USA}

\begin{abstract}
We investigate recent claims that brane inflation solves the overshoot  
problem through a combination of microphysical restrictions on the  
phase space of initial conditions and the existence of the Dirac-Born-Infeld attractor in regimes where the slow-roll attractor does not apply.
Carrying out a comprehensive analysis of the parameter  
space allowed by the latest advances in brane inflation model-building, 
we find that these restrictions are insufficient to solve  
the overshoot problem. The vast majority of the phase space of initial conditions is
still dominated by overshoot trajectories. We present an analytic proof that 
the brane-inflationary attractor must be close to the slow-roll limit, and update the predictions for  
observables such as non-Gaussianity, cosmic string tension and tensor  
modes.
\end{abstract}
\maketitle

\tableofcontents

\section{Introduction}\label{sec:introduction}

The inflationary paradigm postulates a period of 
nearly exponential expansion in the early Universe. This  
ameliorates various initial conditions problems of the classical hot big bang model 
and explains the smoothness of the Universe when averaged on large scales. Furthermore, according to inflation, initial inhomogeneities originated as quantum fluctuations in 
the inflaton field and the metric, sourcing scalar 
\cite{Hawking:1982,Guth:1982, Starobinsky:1982, Bardeen:1983} and tensor 
\cite{Grishchuk:1975, Starobinsky:1979} perturbations which are adiabatic, scale-invariant and almost Gaussian.
\cite{Bardeen:1983}. These broad-brush predictions of inflation are consistent with the latest observations \cite{Komatsu:2008hk, Peiris:2006ug, Kinney:2006qm, Peiris:2006sj, Martin:2006rs, Alabidi:2006qa, Lesgourgues:2007gp, Peiris:2008be, Alabidi:2008ej, Kinney:2008wy, Baumann:2008aq}. 

However, while acknowledging the substantial successes of inflation as a phenomenological model of the early Universe, several theoretical challenges remain.
Chief among these are the lack of a single compelling particle physics motivation for inflation and an understanding of the initial conditions leading to successful inflation.
Attempts have recently been made to address both of
these challenges within the context of string theory (see Refs.~\cite{BMReview, Linde:2006, Cline:2006, Burgess:2007, Tye:2008, Kallosh:2008, McAllister:2008} for recent reviews). In this paper, we shall investigate `warped brane inflation' \cite{KKLMMT}, 
one of the most well-developed string theory models of inflation. Here the inflaton field is identified with
the position of a D-brane in a warped higher-dimensional space, with effective four-dimensional dynamics governed by 
the Dirac-Born-Infeld (DBI) action, and an inflaton potential that is in principle completely derivable from fundamental theory \cite{BMReview}. 

For inflation to be insensitive to initial conditions one typically hopes that the
 inflaton field will naturally flow toward an attractor solution \cite{Belinsky:1985,Felder:2001da,Underwood:2008}, on which the inflaton 
slow-rolls and the Universe inflates. However, should the energy of the inflaton be initially kinetic dominated, it is 
possible for the end of inflation to be reached before the solution has approached the attractor. 
The statement that a large initial kinetic energy spoils inflation is often called the {\sl overshoot problem}~\cite{Brustein:1992nk}.
Due in part to somewhat broad limits on the initial kinetic energy 
in models of inflation based on effective field theory, this occurs for a large proportion of possible inflationary 
trajectories. This poses a theoretical problem, because one of the central aims of inflation is to explain the special 
initial conditions required for the hot big bang. For inflationary models suffering from the overshoot problem, it seems like we have simply exchanged one initial conditions problem for 
another.

The overshoot problem seems particularly acute for small-field models of inflation, where the inflationary 
region only occupies a small area of phase space.
The brane inflation models that we discuss in this paper are of the small-field type \cite{Baumann:2006}
and so seem susceptible to a generic overshoot problem.
However, Underwood~\cite{Underwood:2008} recently observed that brane inflation may not suffer from the overshoot problem,
based on a calculation for a specific set of model parameters. 
This was due to a combination of two effects: a restriction on the allowable initial conditions due to microphysical consistency 
conditions, and an increase in the domain of applicability of the attractor trajectory due to the DBI action. 
Since this claim constitutes an important theoretical argument supporting brane inflation, we will check it for the full range
of parameters, including those identified by recent advances in string theory model-building \cite{Baumann:2008, BaumannMaldacena, BaumannExplicit}. 

Our treatment closely parallels the effective field theory analyses of the overshoot problem
that have been performed for most of the classic inflationary models
(\eg~\cite{Belinsky:1985, Piran:1985, Bird:2008}).
Related work on the parameter space for brane inflation models consistent with the latest experimental bounds has appeared in \cite{Bean:2007, Peiris:2007, Bean:2007eh, Gmeiner:2007, Lorenz:2007,Alabidi:2008, Agarwal:2008, Tzirakis:2008, ChenGong:2008, Cline:2008,Cline:2009pu}. 
Here, we add a systematic study of the overshoot problem and the fine-tuning of inflationary initial conditions for brane inflation.

We perform a comprehensive analysis of the entire parameter space, and determine whether, in general,
these models solve the overshoot problem. We find that they do not; almost all possible trajectories are still dominated by overshoots. 
Furthermore, we reiterate two generic predictions of these models already given in Refs.~\cite{Firouzjahi:2005, Bean:2007}: 
a potentially observable level of cosmic strings generated after the end of inflation
and negligible levels of non-Gaussianity. 
Finally, we examine which regions of the parameter space seem to 
require less tuning of the initial conditions, and consider the extent to which microphysical restrictions 
on the phase space of initial conditions would have to tighten before the overshoot problem for brane inflation can be solved.

\section{D-brane Inflation}

\begin{figure*}
\includegraphics[scale=0.4]{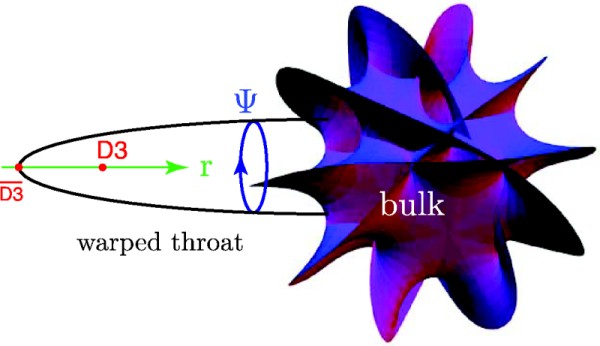}
\caption{Warped D3-brane inflation (figure reproduced from Ref.~\cite{BMReview}).
The 3-branes fill the four-dimensional spacetime and are pointlike in the extra dimensions (shown in this figure). 
The {\it warped throat} has a radial dimension $r$ and a five-dimensional base with coordinates $\Psi$.
At $r_\mathrm{UV}$ the throat attaches to the bulk space. This introduces a cut-off to the throat solution, making the total space finite.
The radial coordinate of the D3-brane in the warped throat plays the role of the inflaton field. 
}
\label{fig:KKLMMT}
\end{figure*}

\subsection{D3-brane motion in warped geometries}

Since the original brane inflation proposal by Dvali and Tye \cite{DvaliTye} many variants of inflation sourced by D-brane motion through a higher-dimensional spacetime have been studied, \eg~\cite{Dvali:2001fw, Alexander:2001ks, Burgess:2001, Burgess:2002, Garciabellido:2002,Jones:2002, Shiu:2001sy, Burgess:2004kv, DeWolfe:2004qx, Iizuka:2004ct, Burgess:2006cb, Pajer:2008uy, Krause:2007jk}. 
In this work we consider a specific, and, arguably, the best understood scenario: D3-brane motion in warped throat geometries \cite{KKLMMT}.

In these scenarios inflation is driven by a mobile D$3$-brane which fills the four-dimensional spacetime $\mathbb{R}^{3,1}$ (with metric $g_{\mu \nu}$), and is pointlike in
the extra-dimensional space $M_6$ (with metric $g_{mn}$).
The ten-dimensional background spacetime is defined by the following warped line element
\begin{equation}
\label{equ:metric}
ds^2 = h^{-1/2}(y) g_{\mu \nu} d x^\mu d x^\nu + h^{1/2}(y) g_{mn} dy^m dy^n\, ,
\end{equation}
where the warp factor $h(y)$ is a function of the coordinates on the internal space $M_6$.
Local approximations to such warped spaces are ubiquitous in the flux compactifications of type IIB string theory\footnote{Readers unfamiliar with this terminology may find a useful Stringlish-to-English dictionary in \cite{Hertzberg:2007ke}.} \cite{DouglasKachru}.
Exact solutions to the supergravity equations may be found \cite{Klebanov:2000} that take the structure of a cone for the extra dimensions
\begin{equation}
g_{mn} dy^m dy^n = d r^2 + r^2 d s^2_{X_5}\, ,
\end{equation}
where $r$ parametrizes the radial direction and $X_5$ is a certain base manifold, for example,
a 5-sphere, $S^5$, or the Einstein space $T^{1,1}$ (arising as  
the limit of the Klebanov-Strassler geometry). These have Vol($S^5$) $= \pi^3$
and Vol($T^{1,1}$)$=\frac{16}{27}\pi^3$, respectively.
The warp factor at large $r$ is approximately
\begin{equation}
\label{equ:AdS}
h(r) \approx \left(\frac{R}{r} \right)^4\, .
\end{equation}
In this limit the spacetime~(\ref{equ:metric}) becomes $AdS_5 \times X_5$.
We define the constant curvature (or warping) scale
\begin{equation}
\frac{R^4}{(\alpha')^2} \equiv 4\pi g_s N \frac{\pi^3}{{\rm Vol}(X_5)}\, ,
\end{equation}
in terms of the string coupling $g_s$, the string length $\ell_s = \sqrt{\alpha'}$, a flux integer $N$ and the dimensionless volume of $X_5$ with unit radius. ${\rm Vol}(X_5)$ generically takes a value of order $\pi^3$.

The motion of a D3-brane in this warped geometry is governed by the DBI action
\begin{equation}
  S = - \int d^4x\sqrt{-g}\left[\frac{\MP^2}{2}\mathcal{R} + P(X,\phi) \right]\, ,
    \label{eq:DBIaction}
\end{equation}  
where $\mathcal{R}$ is the four-dimensional Ricci scalar and 
\begin{equation}
\label{equ:P}
P(X,\phi) =  f(\phi)^{-1}\left(\sqrt{1-2f(\phi) X}-1\right) + V(\phi)\, .
\end{equation}
Here we have defined the canonical inflaton 
$\phi \equiv r \sqrt{T_3}$ in terms of the brane tension
\begin{equation}
		  T_3 = \frac{1}{(2\pi)^3}\frac{1}{\gs}\ms^4\,,
		  \label{eq:branetension}
\end{equation}
where $\ms \equiv 1/\sqrt{\alpha'}$ is the string scale.
The quantity $X \equiv - \frac{1}{2}g^{\mu\nu}\deep_\mu\phi\deep_\nu\phi\approx\frac{1}{2}\dot{\phi}^2$ 
is the corresponding kinetic term, and $V(\phi)$ is the inflaton potential, whose derivation we 
shall sketch in \S\ref{sec:bpotential}. The function $f(\phi) \equiv T_3^{-1} h (\phi)$ is the (rescaled) warp factor 
in terms of the canonical inflaton
\begin{equation}
		  f(\phi) = \frac{\lambda}{\phi^4}\,,
		  \label{eq:warpfactor}
\end{equation}
with $\lambda$ given by
\begin{equation}
		  \lambda \equiv \frac{\pi}{2}\frac{N}{\mathrm{Vol} (X_5)}\,.
		  \label{eq:lambda}
\end{equation}

In the slow-roll limit, $f(\phi) X \ll 1$, the DBI action (\ref{equ:P}) reduces to the canonical slow-roll action
\begin{equation}
P(X,\phi) = - X + V(\phi)\, .
\end{equation}
Slow-roll corresponds to non-relativistic motion of the brane.
The non-slow-roll limit associated with relativistic motion of the brane is 
characterized by a Lorentz-like factor
\begin{equation}
\gamma \equiv [1-f(\phi) \dot \phi^2]^{-1/2}\, .
\label{eq:lorentz}
\end{equation}
For $\gamma \gg 1$ inflation can occur even for relatively steep potentials.
However, large values of $\gamma$ correspond to small values of the effective speed of sound associated with inflaton fluctuations. This leads to large non-Gaussianity in the primordial scalar fluctuations, so that the parameter region of viable relativistic DBI models is significantly constrained by the observational data \cite{Baumann:2006}.

The Hubble parameter, $H$, is given by the sum of potential and kinetic contributions
\begin{equation}
H^2 = \frac{1}{3 \MP^2}\left[V(\phi) + \frac{\gamma-1}{f}\right]\,.
\label{eq:hubble}
\end{equation}

\subsection{Parametrization of the inflaton potential}\label{sec:bpotential}

The computation of the potential $V(\phi)$ for D3-brane inflation has a long and painful history (see Ref.~\cite{BMReview} for a recent review).
Approximations commonly employed in string theory computations have proved to be insufficient when trying to capture all 
effects relevant for inflation.

This challenge is illustrated by the {\it eta problem}: small corrections, $\Delta V(\phi)$, to the 
inflationary potential often lead to large corrections to the second derivative of the potential or the inflaton mass,
causing inflation to end prematurely.
This is an example of the UV sensitivity of inflation where high-energy effects,
 in this case dimension-6 Planck-suppressed corrections, affect the inflationary dynamics.

We now give a brief summary of all relevant contributions to the D3-brane potential 
(more details may be found in \cite{BMReview}).
At zeroth order, the inflaton potential is given by the Coulomb potential between a D3-brane and an anti-D3-brane, 
\begin{eqnarray}
		  \hat{V}_0(\phi) = D \left(1-\frac{3D}{16\pi^2\phi^4}\right) \, , \label{equ:Col}
\end{eqnarray}
where the constant $D$ sets the scale of the brane-antibrane interaction
\begin{equation}
		  D \equiv \frac{2T_3}{h_{\rm IR}}\, .
		  \label{eq:scale}
\end{equation}
Here $h_{\rm IR} \equiv h(\phi_{\rm IR})$ is the warp factor at the tip of the throat (see Fig.~\ref{fig:KKLMMT}).
This warped scale $D$ exponentially suppresses the D3-$\overline{\rm D3}$ interaction, making the potential (\ref{equ:Col}) extraordinarily flat.
In addition there may be distant energy sources 
in the bulk space, which shift the scale of the potential but do not significantly affect the slope or the curvature,
such that
\begin{eqnarray}
		  \hat{V}_0(\phi) = D \left(s-\frac{3D}{16\pi^2\phi^4}\right) \, ,
		  \label{eq:smalletapot}
\end{eqnarray}
where $s \gtrsim 1$ parametrizes the contribution of bulk energy to the cosmological constant during inflation.

 So far, the potential is flat, and eminently suitable for 
supporting a large amount of inflation. In particular, the inflaton mass is small and $\eta \ll 1$. However, we have ignored crucial interactions with moduli-stabilizing ingredients.
Kachru, Kallosh, Linde, Maldacena, McAllister and Trivedi \cite{KKLMMT} identified that the leading moduli-stabilization effect leads to a generic eta problem by
adding a $H^2 \phi^2$ term to the potential, which makes $\eta = \frac{2}{3}$.
However, Ref.~\cite{KKLMMT} also pointed out that further corrections to 
the potential are expected from additional moduli-stabilization effects 
and from coupling of the throat to the bulk geometry \cite{BMReview}. It has recently
become possible to systematically analyze the leading corrections to the brane inflation potential 
\cite{Baumann:2008, BaumannProgress}, using insights from the AdS/CFT correspondence \cite{Maldacena:1997, Gubser:1998, Witten:1998}. 
The potential acquires a correction, $\Delta V$, of the form
\begin{equation}
\label{eq:corr}
\Delta V(\phi, \Psi) =  \sum_\Delta c_\Delta \left( \frac{\phi}{\phi_{\rm UV}}\right)^\Delta f_\Delta(\Psi)\,,
\end{equation}
where $\Psi$ denotes the angular direction, and $\phi_{\rm UV}$ is the value of the inflaton at the large-radius end of the throat.
The scaling dimension $\Delta$ takes a discrete spectrum of values\footnote{The scale of the individual corrections (parametrized by the coefficients $c_\Delta$) is hard to compute without more global information about the compact space and its effects on perturbations at $\phi_{\rm UV}$, the large-radius end of the throat. The constants $c_\Delta$ are therefore treated as free (tunable) parameters.}  \cite{Baumann:2008, BaumannProgress}:
\begin{equation}
\Delta = 1 \ ,\ \frac{3}{2} \ , \ 2 \ ,\ \frac{5}{2} \ , \ \dots
\end{equation} 
Sufficiently deep inside the throat, $\phi \ll \phi_{\rm UV}$ and so only a few of the lowest $\Delta$'s will be relevant.
Furthermore, symmetries can project out some of the $\Delta$ modes.
Notice that the $\Delta =1$ mode does not affect $\eta$, but only corrects the slope of the potential. 
We will absorb it into the small-eta contribution of the potential by defining
\begin{equation}
		  V_0(\phi) \equiv  \hat{V}_0(\phi) + a_1 f_1(\Psi) \frac{\phi}{\MP}\, .  
\end{equation}
In general, the potential implied by (\ref{eq:corr}) has a highly nontrivial angular dependence given by the functions $f_\Delta(\Psi)$.
We will not solve the angular problem in full generality in this work, but study various approximate solutions that we believe capture the qualitative features.
Only in the case when a single $\Delta$ mode is dominant has the effective radial potential been computed analytically
\cite{Baumann:2008}. 
In that case, the resulting potential is
\begin{equation}
  V(\phi) =  V_{0}(\phi) 
  +  \MP^2H^2\left[\left(\frac{\phi}{\MP}\right)^2-a_\Delta\left(\frac{\phi}{\MP}\right)^\Delta\right]\, .
  \label{eq:baumannpotential0}
\end{equation}

The minus sign in Eq.~(\ref{eq:baumannpotential0}), which is crucial for achieving the correct phenomenology, arises because $f_\Delta$ 
is a harmonic function. Any nontrivial harmonic function must have a negative region; thus, if one angular 
mode dominates, minimizing over the angular directions will lead to negative $f_\Delta$. This argument is only 
valid when a single mode of $\Delta$ dominates, hence we consider only that case and small perturbations thereof. 

We believe this approximation to be valid in the inflationary region (in particular, during the last 60 $e$-folds of expansion), but for large values of the inflaton field higher angular modes may become relevant and complicate the angular motion.
When considering the overshoot problem, however, the evolution in this region will usually be dominated by kinetic energy, and hence largely insensitive to the detailed shape of the potential.

Consider, for example, the case where the inflaton potential is determined by the combined effect of the $\phi^{3/2}$, the $\phi^2$, and possible higher $\phi^\Delta$ corrections. Importantly, the two correction terms have different angular dependencies
\begin{equation}
\Delta V(\phi,\Psi) =  a_{3/2} \left(\frac{\phi}{\MP}\right)^{3/2} f_{3/2}(\Psi)  + a_{2} \left(\frac{\phi}{\MP}\right)^{2} f_{2}(\Psi)  \, .
\end{equation}
At some critical value of the field, $\phi_{\rm c}$, (depending on the coefficients $a_{3/2}$ and $a_2$) both terms will be equally important. For $\phi < \phi_{\rm c}$, the $\phi^{3/2}$ term dominates and the potential has a unique angular minimum. For $\phi > \phi_{\rm c}$, the $\phi^2$ term dominates and the location of the angular minimum changes. For a critical region near $\phi_{\rm c}$, both the $\phi^{3/2}$ term and the $\phi^2$ term are dynamically important, so the exact radial motion is hard to compute in a simple way.
In this paper, we will make the simplifying assumption that a single mode $\phi^\Delta$ determines the potential shape throughout the throat region. We know this to be invalid for large values of $\phi$, but since the dynamics in this regime will usually be dominated by kinetic energy, we do not expect a distortion of the potential shape beyond the inflationary region to significantly affect our overshoot analysis.

The corrections to the D3-brane potential were computed in Ref.  
\cite{Baumann:2008} in the `probe approximation', i.e. the most  
general UV perturbations to the throat supergravity solution are  
parametrized and the D3-brane is then used as a probe of that  
perturbed geometry.
It is true that in the relativistic limit the brane kinetic energy can  
lead to additional corrections to the throat geometry.
It is quite possible that these corrections again fall into the  
classification of Ref. \cite{Baumann:2008}, but the full nonlinear  
problem has not been studied and is beyond the scope of the present  
paper.

In any case, once the dynamics has become potential-dominated, our  
branes always slow-roll (see the Appendix). The potential corrections for a
slow-rolling (nonrelativistic) brane derived in Ref. \cite{Baumann:2008} are therefore sufficient.
Furthermore, in the relativistic limit the dynamics is insensitive to  
the form of the potential, but the (corrected) shape of the warp factor would be important.

\vskip 6pt
Two qualitatively different scenarios can now arise, depending on whether the fractional $\Delta =\frac{3}{2}$ mode is part of the spectrum or is projected out.

\vskip 6pt
\noindent
{\sl Quadratic models}.
If the $\Delta = \frac{3}{2}$ mode is projected out of the spectrum, the effective radial potential is
\begin{equation}
V(\phi) = V_0(\phi) + \beta H^2 \phi^2\,. 
\end{equation}
The phenomenology of these types of potentials was first studied analytically by \cite{KKLMMT} and \cite{Firouzjahi:2005},
and numerically by \cite{ShanderaTye}.

\vskip 6pt 
\noindent
{\sl Inflection point models}.
If the fractional mode $\Delta = \frac{3}{2}$ is present, it leads to inflection point models \cite{Baumann:2008, BaumannExplicit, Baumann:2007np, Krause:2007jk}.
If $\phi^{3/2}$ was the only correction term, the effective radial potential would be \cite{Baumann:2008}
\begin{equation}
\label{equ:inf1}
  V(\phi) =  V_{0}(\phi) 
  +  \MP^2H^2\left[\left(\frac{\phi}{\MP}\right)^2-a_{3/2} \left(\frac{\phi}{\MP}\right)^{3/2}\right]\,. 
\end{equation}
We also allow a small $\phi^2$ term to slightly perturb the potential in the inflationary region
\begin{equation}
  V(\phi) =  V_{0}(\phi) 
  +  \MP^2H^2\left[\beta \left(\frac{\phi}{\MP}\right)^2-a_{3/2}\left(\frac{\phi}{\MP}\right)^{3/2}\right]\, ,
  \label{equ:inf2}
\end{equation}
although this does not lead to qualitative changes in our results.
If $\beta \sim \mathit{O}(1)$ then the potential (\ref{equ:inf2}) may be viewed as a small perturbation of the inflection point model  of Eq.~(\ref{equ:inf1})  \cite{Baumann:2008, BaumannExplicit}.

\vskip 6pt
To simplify calculations, we now set $\MP=1$ and replace $H(\phi)$ by its value at $\phi_{\rm UV}$ \begin{equation}
  H^2_\mathrm{UV} \equiv H^2(\phi_{\rm UV}) \approx \frac{V_\mathrm{UV}}{3} = \frac{Ds}{3} + \mathcal{O}(Ds\phi^2)\,.
  \label{eq:HUVapprox}
\end{equation}
This leads to an error in $V(\phi) $ of $\mathcal{O}(D\phi^4)$, the same as the overall uncertainty in 
the potential, which can be removed by a small redefinition of $a_\Delta$. 

Thus the quadratic case is
\begin{equation}
  V(\phi) = Ds\left(1-\frac{C D}{\phi^4} + \alpha\phi + \frac{\beta\phi^2}{3}\right)\,,
  \label{eq:phi2case}
\end{equation}
and the inflection point case is
\begin{equation}
  V(\phi) = Ds\left(1-\frac{C D}{\phi^4} + \alpha\phi  +\frac{\beta}{3}\left[\phi^2 - a_{3/2}\phi^{3/2}\right]\right)\, ,
  \label{eq:phi32case}
\end{equation}
where for convenience we have defined 
\begin{equation}
		  C \equiv \frac{3}{16\pi^2} \frac{1}{s}\, .
\end{equation}

{\sl Parameters of the potential.}\label{sec:parameters}
To be sure that the argument regarding the minimization of the angular function is valid, and that the 
sign of the leading coefficient is such that it can cancel the inflaton mass term, one single power of $\phi$ must be 
dominant in Eq.~(\ref{eq:corr}). The linear term (which cannot solve the eta problem) must therefore be smaller than 
the leading $a_\Delta$ term, {\sl i.e.}
\begin{equation}
		  \alpha\phi < \frac{a_\Delta \phi^\Delta}{2}
		  \label{eq:alphaprior}
\end{equation}
throughout inflation. We do not know the exact location of the tip, so we impose this at the end of inflation,
when $\epsilon=1$, ensuring that it is the case for all larger $\phi$. 
For the inflection point case, $a_2$ must be small compared to $a_{3/2}$ 
(although it need not be positive), and so we also impose $0.9 < \beta < 1.1$. 

\subsection{Initial conditions and overshoot}\label{sec:initconds}

The basic objective of this paper is to examine the overshoot problem for models of brane inflation. To quantify the effect of an initially large kinetic energy, we define the conjugate momentum
\begin{equation}
\Pi \equiv \frac{\deep \mathcal{L}}{\deep \dot{\phi}}  = \frac{\dot{\phi}}{(1-f(\phi)\dot{\phi}^2)^{1/2}} = \gamma \dot \phi\, ,
\label{eq:conjugatemom}
\end{equation}
where ${\cal L}$ is the Lagrangian density corresponding to the DBI action (\ref{equ:P}).
We can redefine the Lorentz factor (\ref{eq:lorentz}) in terms of the conjugate momentum (\ref{eq:conjugatemom})
\begin{equation}
\gamma \equiv \sqrt{1+f(\phi)\Pi^2} \,.
\end{equation}

We wish to analyze the effect of the initial field value $\phi_\mathrm{i}$ and a possibly large initial momentum 
$\Pi_\mathrm{i}$ on classical inflationary trajectories in a flat, homogeneous Universe. We shall apply a 
criterion similar to that applied to, for example, chaotic inflation in Refs.~\cite{Belinsky:1985, Piran:1985}. 
The criterion $\Ri$ for a given potential is the 
proportion of the total phase space which leads to successful inflation \cite{Bird:2008}, and is defined by

\begin{equation}
\Ri \equiv \frac{\int{F(\phi_\mathrm{i},\Pi_\mathrm{i})d\phi_\mathrm{i} d\Pi_\mathrm{i}}}{\int d\phi_\mathrm{i} d\Pi_\mathrm{i}}\,,
\end{equation}
where $F(\phi_\mathrm{i},\Pi_\mathrm{i})$ is an indicator function unity where ($\phi_\mathrm{i}$, $\Pi_\mathrm{i}$)
leads to successful inflation, and zero otherwise. 
We define a successful trajectory to be one which satisfies all microphysical and observational constraints (see below). 
Minimal fine-tuning of initial conditions corresponds to $\Ri=1$, and maximal fine-tuning 
to $\Ri=0$. Here we have used the Hamiltonian measure of Ref.~\cite{Gibbons:1987}, canceling a factor of $a^3(\phi_\mathrm{i})$ which is the same for all parameters. Our main results do not depend on the details of the measure, but on the upper bound for $\Pi$ derived below. Hence our conclusions should be largely unaltered should an alternative measure be used.

This is a well-defined criterion, because there are microphysical 
constraints on the allowable initial conditions that make the total phase space finite.\footnote{This is to be contrasted with the corresponding effective field theory analysis, where ad-hoc limits on the phase space need to be introduced.} A lower limit is set upon 
$\phi_\mathrm{i}$ because the branes must initially be separated by at least a string length, so that inflation does not end immediately. We also assume that
$\Pi_\mathrm{i} \geq 0$, so that $\phi$ evolves monotonically. We shall now 
review briefly two other important constraints on the phase space for brane inflation, arising from microscopic 
consistency requirements, which bound the initial conditions phase space from above.

\vskip 6pt
{\sl Field range constraint}.
A restriction on the maximal field range of the inflaton for brane inflation models was derived in Ref.~\cite{Baumann:2006};
heuristically, this bound arises because the inflaton now has a geometrical interpretation, and cannot move 
a distance that is larger than the size of the compactified space. At the same time the four-dimensional Planck mass (the unit in which the field excursion of the canonical inflaton field is measured) also scales with the volume of the compact space. By considering only the volume of 
the space inside the warped throat (omitting any motion within the bulk) one obtains an 
upper bound on the inflaton field range for a given warp factor. 
For the AdS warp factor considered in this work, the field range bound ultimately has the following simple form \cite{Baumann:2006}:
\begin{equation}
  \left(\frac{\Delta\phi}{\MP}\right)^2 <\frac{4}{N}\,.
  \label{eq:bmbound3}
\end{equation}
For the throat to be warped we require that $N\geq 1$, and the validity of the supergravity 
description requires that $N \gg 1$. We shall take $N=10$ to be the lower limit (the exact number 
makes little difference to the results), and thus obtain a lower bound on the warping and an upper bound 
on the field range. We see that in the regime of control the field range is always sub-Planckian.

\vskip 6pt
{\sl Backreaction constraint}. 
To constrain the initial inflaton momentum we consider the requirement that the D3-brane not contain enough energy to trigger the production of significant amounts of Kaluza-Klein 
modes, which would back-react on the geometry. This is required for the effective four-dimensional description to be valid, 
and implies that the Hubble parameter is everywhere
less than the local warped string scale \cite{Underwood:2008, Frey:2006, Becker:2007, Kobayashi:2008}. At the tip, where the warped string scale is smallest, we impose the condition
\begin{equation}
		  H \leq \frac{\ms}{h^{1/4}_{\rm IR}}\,.
  \label{eq:KKbound}
\end{equation}
Using Eq.~(\ref{eq:scale}), we get
\begin{equation}
		  H < \sqrt{2}\pi^{3/4}\, D^{1/4}\gs^{1/4}\, .
\end{equation}
Finally, this can be written as a function of the conjugate momentum 
\begin{equation}
		  \frac{\Pi}{\sqrt{2(\gamma+1)}} \ \ < \ \ \sqrt{6\pi^{3/2}\,D^{1/2}\gs^{1/2} - V(\phi)}\,.
  \label{eq:KKbound2}
\end{equation}
For concreteness, we shall take $\gs = 0.01$.

\section{Dynamics and Observables}

In the preceding section, we examined the model of D-brane inflation, and the theoretical consistency conditions that 
attach to it. 
We shall now discuss the evolution of the inflaton, define observables, and briefly review the relevant observational constraints, completing the definition of successfully inflating trajectories.

\subsection{Equations of motion}\label{sec:eqsofmotion}

The inflationary dynamics are controlled by the DBI action (\ref{eq:DBIaction}).
For our time coordinate we shall use the number of $e$-folds of inflation,
\begin{equation}
  dN = H dt\,.
\end{equation}
Using the canonical momentum of Eq.~(\ref{eq:conjugatemom}), we can write the equations of motion in the following form
\begin{eqnarray}
 \frac{d\phi}{dN}  &=& \frac{\Pi}{H\gamma}, \\
\frac{d\Pi}{dN} &=& -3\Pi - \frac{V'(\phi)}{H} + \frac{f'(\phi)}{2H\gamma}\left(\frac{\gamma-1}{f}\right)^2\,,
\end{eqnarray}
where a prime denotes a derivative with respect to $\phi$.
Our conventions are chosen such that $\phi$ decreases towards the end of inflation, in agreement 
with the sign convention used, for example, in Ref.~\cite{Peiris:2007}. We solve this system of equations numerically
in the following analysis.

\subsection{Cosmological observables}

{\sl Power spectra}.
We shall calculate observables to first order in the Hubble slow-variation parameters
\begin{eqnarray}
\epsilon &=& \frac{2}{\gamma}\left(\frac{H'}{H}\right)^2\,, \\
\eta &=& \frac{2}{\gamma}\left(\frac{H''}{H}\right)\,, \\
\kappa &=& \frac{2}{\gamma}\left(\frac{\gamma' H'}{\gamma H}\right)\,.
\end{eqnarray}
The power spectra of scalar and tensor fluctuations are then given (to first order in the slow-variation parameters) by
\begin{eqnarray}
  \Ps(k) &=& \left.\frac{\gamma}{8\pi^2\MP^2}\frac{H^2}{\epsilon}\right|_{k=a H\gamma}\,, \\
  \Pt(k) &=& \left.\frac{2}{\pi^2}\frac{H^2}{\MP^2}\right|_{k=a H}\,. 
  \label{eq:powerspec}
\end{eqnarray}
Notice that scalar perturbations freeze out at $k=a H\gamma$, when exiting the sound horizon,  
while tensor modes freeze out at $k=a H$. We can parametrize small deviations 
from scale-invariance using the scalar spectral index $\ns$, and measure the magnitude of 
gravitational waves using the tensor-scalar ratio $r$
\begin{eqnarray}
  \ns-1 &=& \frac{d\Ps}{d \ln k} = 2\eta-4\epsilon-2\kappa\,, \\
  r &=&  \quad \frac{\Pt}{\Ps} \quad\approx \quad\frac{16}{\gamma}\epsilon\,.  \label{equ:rr}
\end{eqnarray}
Because $\Pt$, the tensor power spectrum, is evaluated at freeze-out for the tensor modes $k=a H$, 
the second equality in (\ref{equ:rr}) holds only in the slow-roll regime, where $\gamma \approx 1$. When we calculate 
$\ns$ and $r$, we shall always start the inflaton on the DBI attractor trajectory, as described by 
Eq.~(\ref{eq:inflatattrsol}) in Appendix \ref{sec:slowroll}. 
When we 
consider the fine-tuning of the initial conditions we shall quantify the effect of starting the inflaton off the attractor. 
In the following analysis, we assume that these observables are measured at the typical physical scales 
corresponding to cosmic microwave background (CMB) measurements.

\subsection{Observable constraints}

The most important observational constraints on inflation are provided by the CMB. In order to predict the 
observables for brane inflation, 
we must compute the evolution of the inflaton, find the end of inflation, and then 
work backward to determine the $e$-folds corresponding to the CMB scales, taking reheating into account.

\vskip 6pt
{\sl The end of inflation.}
Within a string length of the tip of the throat, string modes between the D$3$-brane and the 
anti-D$3$-branes will become tachyonic \cite{KKLMMT}. The resulting instability will annihilate the branes 
and end inflation. However, near the tip, the warp factor deviates from the form of Eq.~(\ref{eq:warpfactor}) 
in a model-dependent way. This makes it difficult to calculate the value of $\phi$ when the tachyonic transition occurs. 
Fortunately, it is not necessary; inflation will end when $\epsilon>1$, so we need simply make sure that this 
happens before the tip of the throat.

While we cannot calculate the value of $\phi$ at the tip exactly, we can find an approximation to it by 
combining Eqs.~(\ref{eq:warpfactor}) and~(\ref{eq:scale}). This is a very conservative upper bound, 
because the true form of the warp factor is much flatter than this, \eg~\cite{Klebanov:2000}. 
We found that this upper bound is generally slightly larger than $\phie$, at which inflation ends via 
$\epsilon=1$, but that the differences between the two quantities do not significantly change the 
observables. Thus, we can 
be reasonably confident that the tip of the throat is not reached before the breakdown of slow-roll.

We encountered a small class of models for which $\gamma$ became very large in the 
region of the tip, preventing $\epsilon>1$. With the AdS warp factor, these models are all ruled out 
by the field range bound. However, since inflation here is likely to continue right up to the tachyonic transition, 
it is possible that model-dependent corrections to the warp factor near the tip could weaken the field range bound 
sufficiently to prevent the model from being ruled out. This could lead to distinctive observational signatures, such 
as primordial black holes, but because $\gamma$ only becomes large near the end of inflation, is unlikely to 
lead to large non-Gaussianity on CMB scales. The calculation of such warp factor corrections is beyond the scope of 
this work. We simply assume the AdS warp factor throughout, and thus consider such models ruled out by the field 
range bound.

\vskip 6pt
{\sl Reheating uncertainty}.
Reheating after brane inflation is very model-dependent \cite{Kofman:2005}.
Here, we make the simplifying assumption of instantaneous reheating.
From the point of view of the potential,
more sophisticated reheating models would simply modify the number of effective $e$-folds needed for inflation 
to solve the horizon problem, thus shifting the part of the potential to which the CMB scales correspond. This is 
roughly equivalent to a small change in the value of $s$ in Eq.~(\ref{eq:smalletapot}). Because we already allow $s$ to
vary, this should not change the results.

\vskip 6pt
{\sl Horizon scales.} 
If we assume instantaneous reheating, a scale $k$ exits the horizon $N$ $e$-folds before the end of inflation; 
\begin{equation}
  N(k) - \log\left[H(N)\right] = 67.9 -\frac{1}{2}\log\left[H(\phie)\right]\,.
  \label{eq:connection}
\end{equation}
Here $\phie$ is the value of the inflaton at the end of inflation. 
We can use Eq.~(\ref{eq:connection}) to deduce the location of the CMB scales in terms of $N$. This depends 
slightly on the potential, but in practice it is usually around $N=58$. In order to solve the horizon problem,
inflation must last at least this long. 

\vskip 6pt
{\sl WMAP normalization}.
The amplitude of the power spectrum has been accurately determined by experiments such as WMAP
\cite{Komatsu:2008hk} to be, at the $68\%$ confidence level, $\Ps=(2.445\pm 0.096)\times 10^{-9}$ at $k=0.002$ Mpc$^{-1}$. We impose the constraint that $\Ps$ is within $1.5\sigma$ of its observed value. 
This essentially fixes the overall scale of the potential, 
$D s$ in Eq.~(\ref{eq:baumannpotential0}). 

The only other measurement currently strong enough to significantly 
constrain the model is that on $\ns$, but this is not imposed because we would like to see the full range of observables predicted by the self-consistent model parameter space.

\subsection{Priors on the potential parameters}

In the quadratic case, $\beta$ lies between zero and one. Since all models are equivalent for a sufficiently small $\beta$, we take $\beta > 10^{-5}$ as a lower limit.

In the inflection point case, a fine-tuning is required in the parameter $a_{3/2}$ to achieve inflation, so we simply 
place broad limits upon this parameter and 
allow the sufficient inflation constraint to set effective limits. 

In order to avoid complications with reheating, we assume that $s$, which measures the contribution to 
the total potential energy from the bulk, does not 
dominate enormously over the potential energy from the brane-antibrane pair. Concretely, we assume $s<10$. 
Table \ref{table:priors} summarizes our priors for the entire parameter space.

\begin{table}[!htp]
\begin{center}
\begin{tabular}{|c|c|c|}
\hline
Parameter & Lower limit & Upper limit \\ 
\hline
$\log\beta$ (quadratic)	& $-5$	& $0$ 	\\
$\beta$ (inflection) & $0.9$		& $1.1$ 	\\							
$a_{3/2}$ 				& $0$			& $1$ 	\\
$\log s$ 						& $0$			& $1$	\\
$\alpha$ 				& $0$ 		& Eq.~(\ref{eq:alphaprior}) \\
$\log\lambda$ 				& $-0.3$\footnote{From Eqs.~(\ref{eq:lambda}) and (\ref{eq:bmbound3}).}& Eq.~(\ref{eq:bmbound3}) \\
$\phi_\mathrm{i}$ 	& $\phie$ 	& Eq.~(\ref{eq:bmbound3}) \\
$\Pi_\mathrm{i}$ 		& $0$ 		& Eq.~(\ref{eq:KKbound2}) \\
\hline    
\end{tabular}
\end{center} 
\caption{Summary of parameter space. Uniform priors were placed on the quantities in the leftmost column.}
\label{table:priors}
\end{table}

\section{Results}\label{sec:results}

\subsection{Observables}\label{sec:observables}

\begin{figure*}
\includegraphics[scale=0.15]{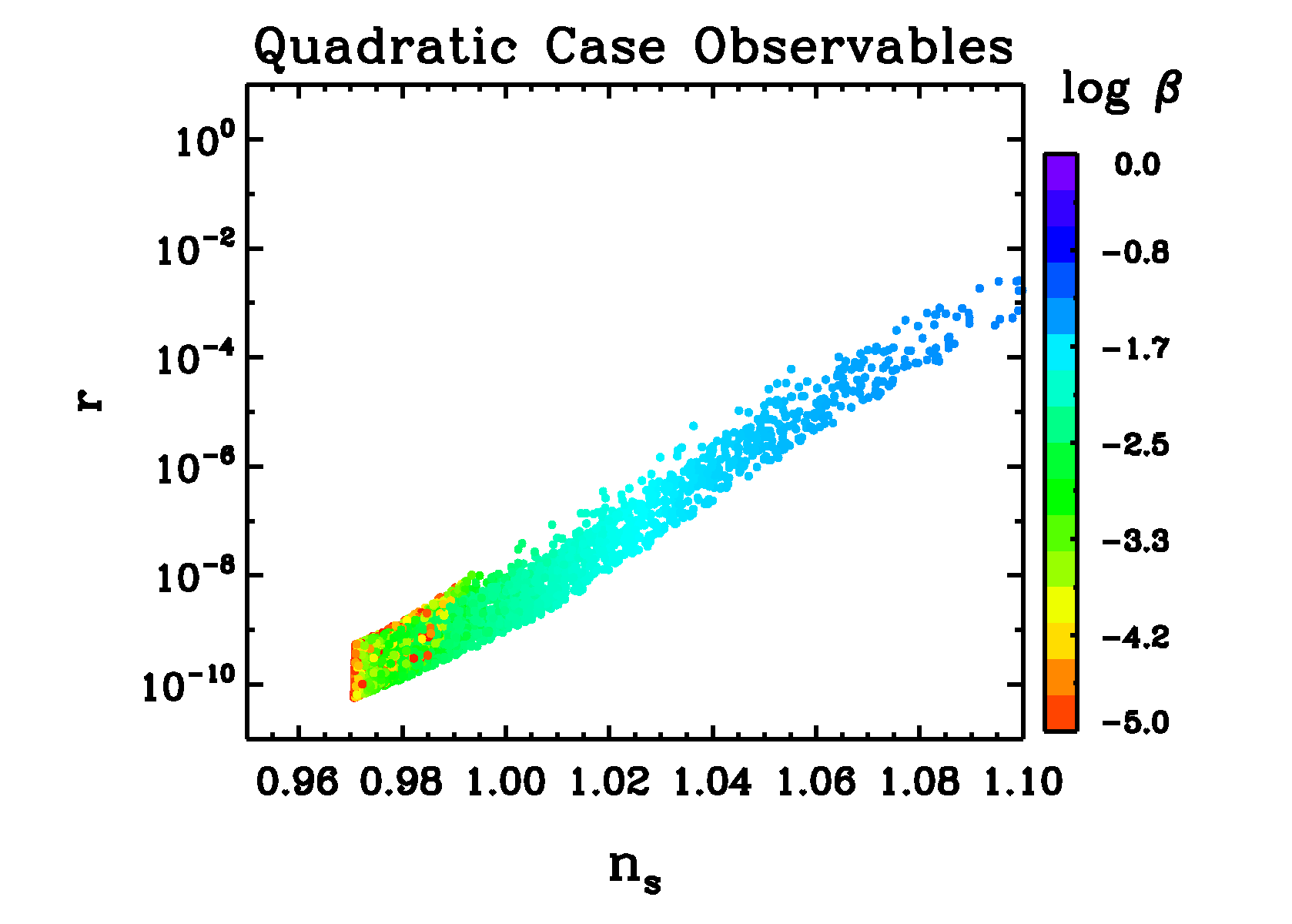}\hfill
\includegraphics[scale=0.15]{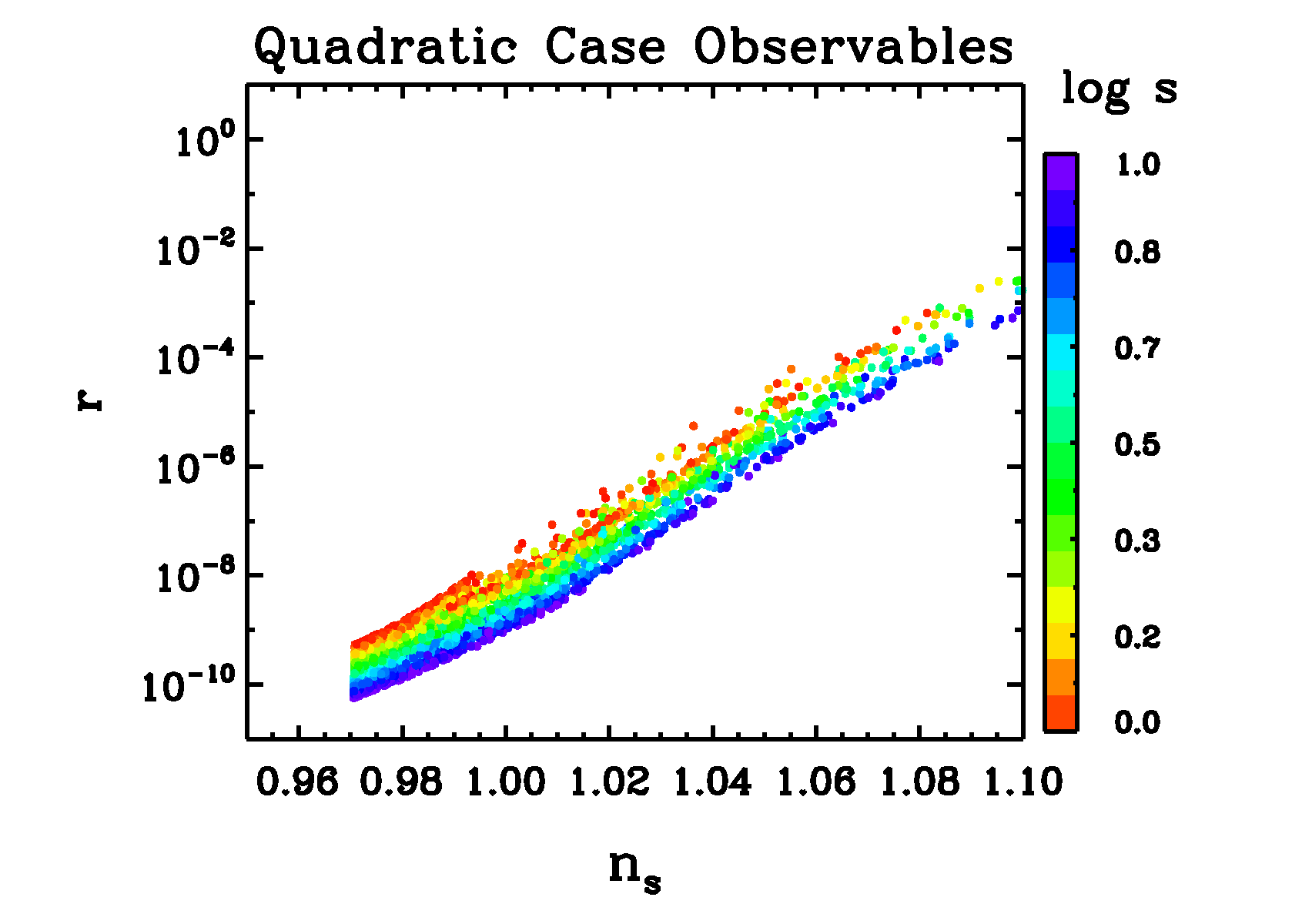}\hfill \\

\caption{The scalar spectral index $\ns$ vs the
tensor-scalar ratio $r$ for the quadratic case. Left: color scale shows $\log\beta$, 
the size of the mass correction to the potential. 
Right: color scale shows $\log s$, the magnitude of the bulk energy in units of the warped string scale at the tip.}
\label{fig:params_all}
\end{figure*}

In this section we consider the cosmological observables, ignoring the overshoot problem for the moment. We shall choose the initial conditions
specifically in order to minimize the amount of overshoot; this implies a large $\phi_\mathrm{i}$, and $\Pi_\mathrm{i}$ chosen so 
that the evolution 
starts on the inflationary attractor. Each dot in Figs.~\ref{fig:params_all} and \ref{fig:initialconditions} 
corresponds to a set of parameters which generated a successful trajectory (as defined in the previous sections) with these 
special initial conditions, and the values of $\ns$ and $r$ plotted for each dot are those for this trajectory. 
Reference~\cite{Bean:2007} performed this 
analysis for the quadratic case, and we reproduce their results where our analyses overlap.

Figure~\ref{fig:params_all} plots the observables $\ns$ and $r$ against values of the potential parameters, 
for the quadratic case. The scatter in $r$ is largely due to variation in the values of the parameter $s$, while
 $\beta$ controls $n_s$. A significant value of $\alpha$  
would boost the tensor-scalar ratio, without affecting $\ns$. Since we are not allowing $\alpha$ 
to dominate the dynamics, it just leads to additional scatter in $r$. The equivalent graphs for the inflection point case 
follow a similar pattern, although $\ns$ is generically much bluer than in the quadratic case, reflecting the increased
difficulty of circumventing the $\eta$ problem. 

In neither case are observable tensor modes generated.
This is consistent with the Lyth bound in combination with the field range limit of Ref.~\cite{Baumann:2006}.

Points with the red tilt favored by WMAP have small $\beta$. Red tilt requires $\eta <0$, 
and, since we always have $\gamma \approx 1$ (see Appendix A), this implies $V''<0$. This implies
\begin{equation}
2\beta\phi <  \frac{20 \, CD}{\phi^6}\,.
\end{equation}
Equation~(\ref{eq:alphaprior}), the limit on $\alpha$, is in terms of $a_2= 1-\beta$, and so a larger 
value of $\alpha$ is allowed for such small $\beta$ models. 
Thus the scatter due to $\alpha$ is more pronounced for parameters giving rise to a red tilt.

When $\beta \lesssim 10^{-5}$, the quadratic term is dynamically unimportant and the potential is 
\begin{equation}
  V(\phi) = Ds\left(1-\frac{C D}{\phi^4}\right)\,.
\end{equation}
Thus the predictions for all models with very small $\beta$ are essentially identical, 
as derived analytically in Appendix C of Ref.~\cite{KKLMMT}.

The Lorentz factor $\gamma$ is very close to unity for all models, quadratic and inflection point. 
We found that it was not possible to generate an observable degree of primordial non-Gaussianity with 
this potential if all constraints are satisfied, in agreement with the results of \cite{Bean:2007}. 
A proof is presented in Appendix \ref{sec:slowroll} that, for the potential (\ref{eq:baumannpotential0})
and the AdS warp factor, the inflationary attractor slow-rolls. 
The proof makes use of the field range bound of Eq.~(\ref{eq:bmbound3}) and significant non-Gaussianity can be generated 
if this restriction is ignored \cite{Bean:2007}. A successful trajectory will in general be close to the attractor for 
most of its evolution, and thus close to the slow-roll limit. We found no trajectories, for any initial conditions, 
which generated observable non-Gaussianity on CMB scales. 

Because all models are slow-roll, the nonstandard kinetic term does not have a significant effect on 
the dynamics, and so the value of $\lambda$ is uncorrelated with $\ns$ and $r$. 

The brane-antibrane annihilation which ends inflation will generate cosmic strings, 
at a dimensionless amplitude given by Ref.~\cite{Firouzjahi:2005}
\begin{equation}
		  G\mu = \sqrt{\frac{1}{32\pi g_s}}\left[\frac{1}{f_{\rm IR}}\right]^{1/2}\,.
  \label{eq:cosmicstrten}
\end{equation}
Here $g_s<1 $ is the string coupling, and $f_{\rm IR}$ is the (rescaled) warp factor at the tip of the throat. Taking a 
fiducial value of $\gs=0.01$ and using Eq.~(\ref{eq:scale}), we can calculate $G\mu$.
If we assume a red spectral tilt, we have $5 \times 10^{-9} < G\mu < 2\times 10^{-8}$ for the quadratic case, and 
$7 \times 10^{-11}  < G \mu < 2\times 10^{-9}$ for the inflection point case. A significantly larger value of $s$ would lower 
these values. 
Models with a redder spectral tilt need a smaller warp factor at the tip to match the CMB amplitude, and thus lead to a lower value of $G\mu$. 
Where they overlap, these results are in agreement with Refs. \cite{Firouzjahi:2005, Bean:2007}. 

The current best observational bounds on cosmic strings come from searching for their signatures in the CMB \cite{Lo:2005}, 
and from nondetections of the gravitational radiation that would result from the decay of string networks.
Millisecond pulsar timing surveys provide the best constraints of the latter type, but are dependent 
on assumptions about the overall level of background gravitational waves. Constraints vary between 
$G\mu < 10^{-6}$ and $G\mu < 10^{-7}$ \cite{Battye:2006}. 

This is insufficient to dramatically limit the model, but future weak lensing surveys or CMB experiments could 
improve constraints to $G\mu < 10^{-9}$ \cite{Mack:2007, Planck, Fraisse:2008}, while the LISA satellite could 
improve upon that by several orders of magnitude \cite{Depies:2007}.

\subsection{Implications for the overshoot problem}

We shall now relax the special choice of initial conditions in the last section, and consider the full $(\phi_\mathrm{i}, \Pi_\mathrm{i})$ 
plane compatible with the initial conditions bounds of \S\ref{sec:initconds}, checking each successful set of parameters from the previous section for trajectories suffering from overshoot. 

We find that, for every set of parameters allowed by the constraints, $\Ri < 5 \times 10^{-3}$. That is, more than $99.5\%$ of initial 
conditions lead to trajectories suffering from overshoot.
This was because, in contrast to the specific case considered by Ref.~\cite{Underwood:2008}, the bound on the initial kinetic energy was 
insufficiently restrictive to solve the overshoot problem. The other effect identified in Ref.~\cite{Underwood:2008}, the increased 
extent of the inflationary attractor, 
and consequent increase in $e$-folds, was also found to be ineffective. Both effects required a much stronger warping than was allowed 
by the field range bound and the observational constraints. 

We have determined that the entire ``successful" parameter space suffers from overshoot, but would like to know which parameter combinations suffer more than others. 
Unfortunately, the fact that overshoot trajectories dominate makes this difficult, so we introduce an additional, ad-hoc,
kinetic bound of $\Pi_\mathrm{i}\leq 10^{-7}\MP^2$ purely as a presentation aid for the following results. We call the analogous rescaled initial conditions fine-tuning criterion 
$\Rit$. The bound was chosen so that the least fine-tuned quadratic 
model had $\Rit$ of about unity, and then applied to both the quadratic and inflection point cases. 
We calculated $\Rit$ to an accuracy of about $1\%$, using a grid-refinement method.

\begin{figure*}
\includegraphics[scale=0.15]{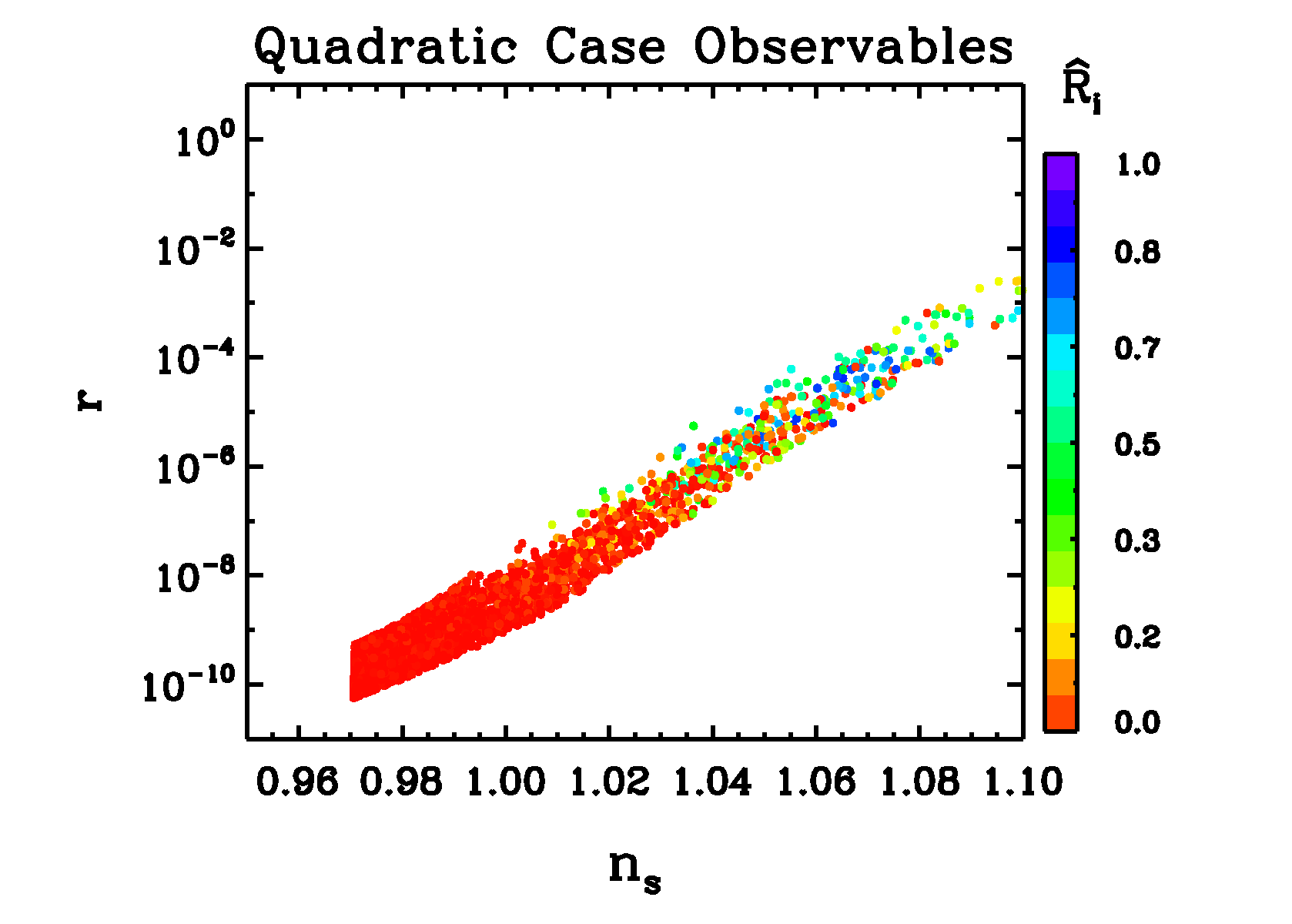}\hfill
\includegraphics[scale=0.15]{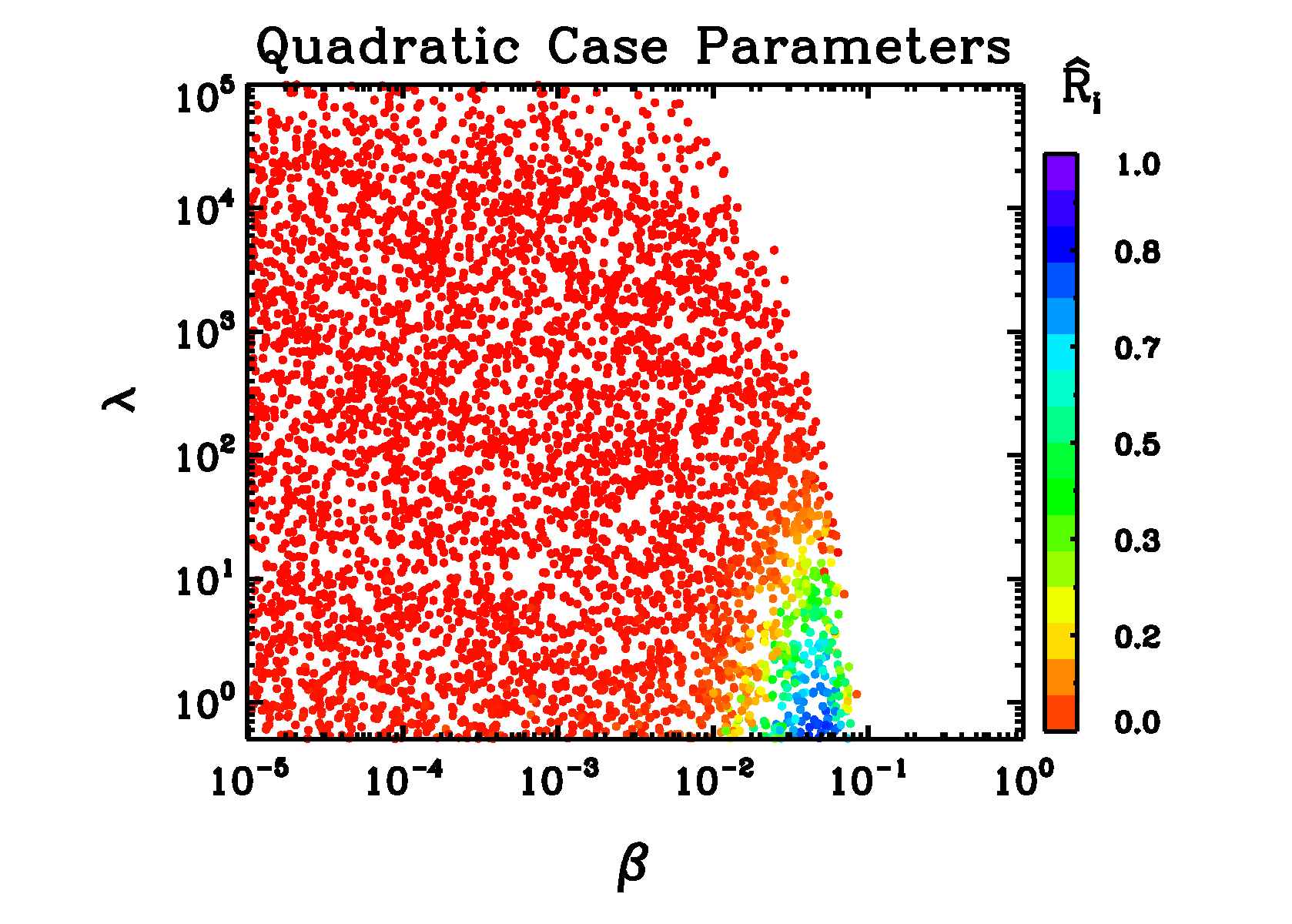}\hfill \\
\includegraphics[scale=0.15]{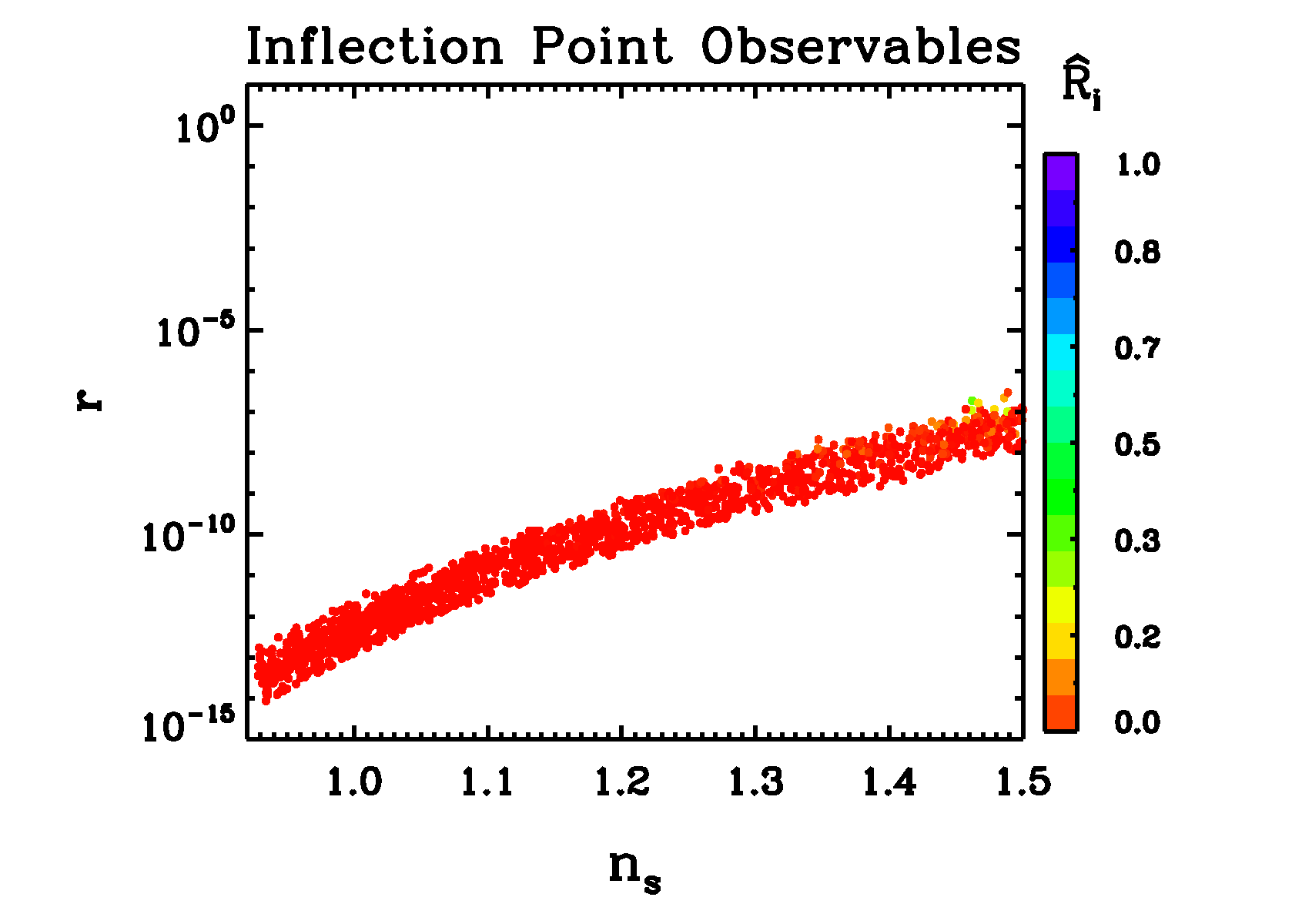}\hfill
\includegraphics[scale=0.15]{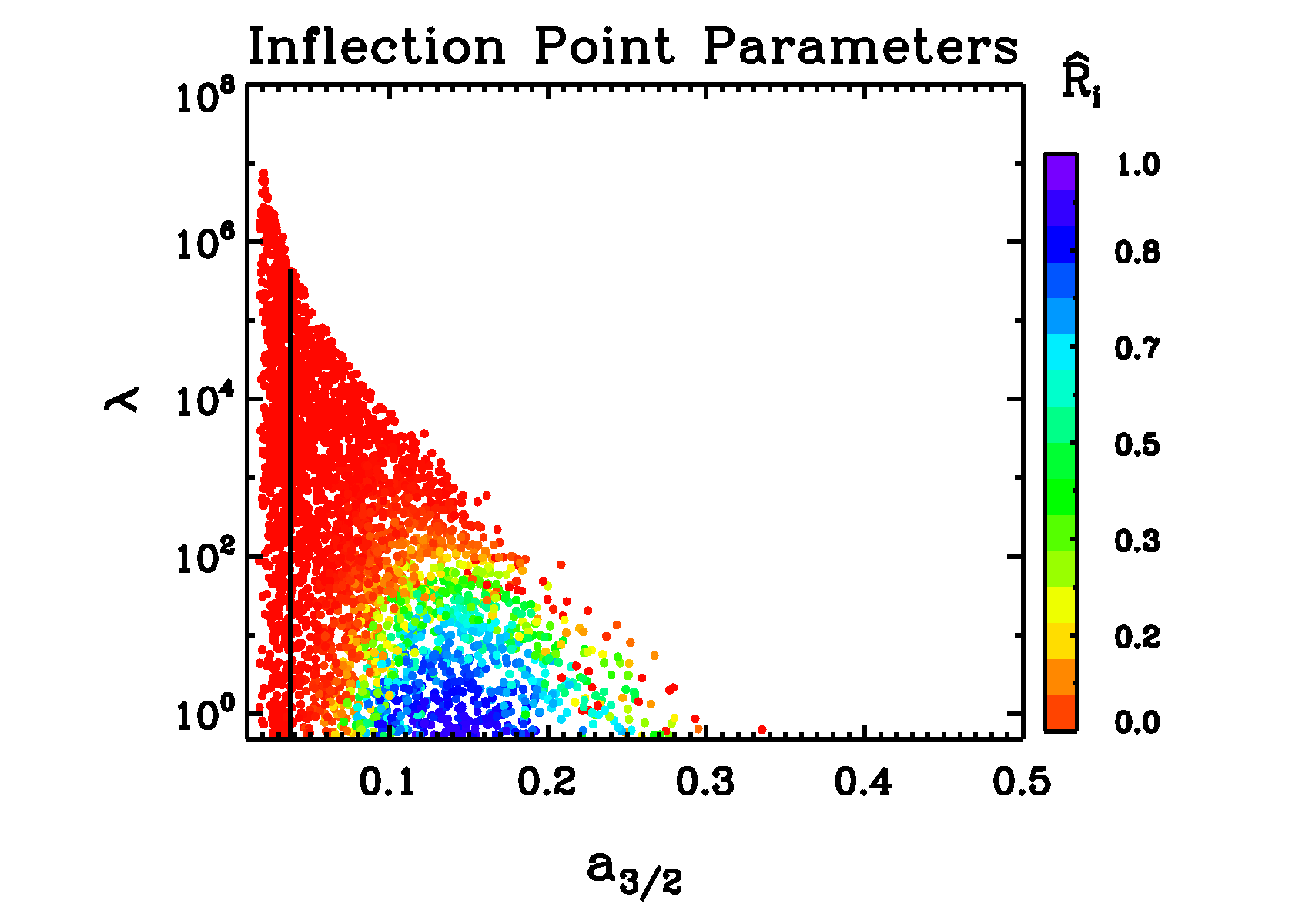} \hfill \\

\caption{The color scale shows the rescaled initial conditions criterion $\Rit$.
Top: results for the quadratic case. Bottom: results for the inflection point case.
Left: scalar spectral index $\ns$ plotted against the
tensor-scalar ratio $r$. Right: strength of the warping scale $\lambda$ plotted against the
magnitude of the mass term ($\beta$ for the quadratic case, $a_{3/2}$ for the inflection point case).} 
\label{fig:initialconditions}
\end{figure*}

Figure~\ref{fig:initialconditions} shows the results for $\Rit$. 
For observationally relevant values of $\ns$, the 
quadratic case suffers less from overshoot than the inflection point case. 
Points with $\ns< 1.0 $ always have $\Rit \approx 0$. These potentials have small $\beta$ or $a_{3/2}$, and 
so are very flat in the region of interest. Because of this, they
must dissipate more kinetic energy before becoming potential--dominated, making them more susceptible to overshoot. 
The flatter the potential, the less field variation is required to generate sufficient inflation. 
Thus the flatter models, which also have $\ns<1$, can have a larger warping scale than those with $\ns>1$,
without violating Eq.~(\ref{eq:bmbound3}). 

For blue tilted models, the most important parameter controlling $\Rit$ is $\lambda$, the warping scale. 
The warping scale is largely uncorrelated with the observables, which leads to
the slightly chaotic appearance of the top-left plot of Fig.~\ref{fig:initialconditions}. 
It is apparent from the right-hand plots of Fig.~\ref{fig:initialconditions}, 
which show $\Rit$ as a function of the potential parameters, that
a high value of $\lambda$ can make it harder to achieve inflation, even though this means a smaller 
initial conditions phase space. The effect of increasing $\lambda$ is to tighten the 
field range bound, Eq.~(\ref{eq:bmbound3}). A tighter field range bound
allows the inflaton less space to dissipate excess kinetic energy, making overshoot more severe.

\vfil

\subsection{Successful inflection point inflation}

The generic behavior of the potential in the inflection 
point case involves either super-Planckian field variations (disallowed by the field range bound), when the $a_{3/2}$ correction is too small, 
or else a metastable minimum of the potential when it is too large.
Successful inflation can only occur in a carefully balanced intermediate regime between these two effects,
when the minimum has become an inflection point. This requires a fine-tuning of the model parameters and the initial conditions, which we shall now quantify. 
We shall assume $\alpha=0$ for simplicity. In order to have an inflection point, we need
\begin{equation}
		  5\times 10^{-3} < a_{3/2} < 1.5\,.
  \label{eq:finetuned1}
\end{equation}
If $a_{3/2} > 1.5$, a minimum always occurs for $\phi>\MP$. This could 
be transformed into an inflection point potential if $D$ was sufficiently large, but inflation 
would then occur in the vicinity of the former minimum, where \\ 
$\phi>\MP$, which is always outside the throat.
If $a_{3/2}< 10^{-3}$, the potential is insufficiently flat and the field variation is super-Planckian, violating the field range bound. 

Once we have chosen an $a_{3/2}$, we need to pick $D$ carefully.
It is helpful to reparametrize $C D$ in terms of $a_{3/2}$
\begin{equation}
		  C D \equiv w \left(\frac{3a_{3/2}}{4}\right)^{12}\,.
  \label{eq:finetuned2}
\end{equation}
To get inflation we need $5.3\times 10^{-3} < w < 6\times 10^{-3}$. Once we combine this limit with
the observed amplitude of the power spectrum, we need: 
\begin{equation}
		  0.01 < a_{3/2} < 0.4\,.
  \label{eq:finetuned3}
\end{equation}
The bottom-right plot of Fig.~\ref{fig:initialconditions} shows this fine-tuning graphically, and the further fine-tuning 
that results when we also require that $\ns< 1.1$. 

A nonzero $\alpha$ can weaken this tuning, and increase the generic value of $r$, but it does not significantly 
change the observables, unlike in the quadratic case.

\section{Conclusions}

Recently, Ref.~\cite{Underwood:2008} proposed an intriguing solution to the overshoot problem for brane inflation.\footnote{Other creative proposals for solving the overshoot problem of brane inflation have appeared in \cite{Itzhaki:2007nk, Itzhaki:2008ih, Spalinski:2009rq}.} 
Theoretical limits on allowed initial conditions shrunk the volume of 
phase space, removing most of the trajectories leading to overshoot, while the existence of the DBI attractor 
extended to regions where the slow-roll attractor did not, thus allowing more inflation to take place. 
However, Ref.~\cite{Underwood:2008} studied the dynamics only for a very specific choice of parameters in the inflationary potential and the warp factor, 
so it remained unclear if these two effects were 
sufficient to solve the overshoot problem in general. 

In this paper we performed a comprehensive scan of the parameter space allowed by the latest advances in brane inflation model-building \cite{Baumann:2008, BaumannMaldacena, BaumannExplicit}. 
For each set of parameters we first worked out the observables for a particular set of initial conditions, 
chosen so that they were least likely to suffer from the overshoot problem. If this trajectory satisfied all microphysical and observational constraints, 
we quantified the degree of overshoot by calculating the proportion of the
homogeneous phase space of initial conditions leading to successful inflation. We 
found that in all cases this was less than $5\times 10^{-3}$; that is, less than $0.5\%$ of 
the phase space volume of initial conditions led to successful trajectories. Therefore, brane inflation does not significantly alleviate the overshoot problem.

The qualitative differences in our conclusions compared to Ref.~\cite{Underwood:2008} are easy to understand:
Most importantly, the parameters chosen by Ref.~\cite{Underwood:2008} violate the field range bound (\ref{eq:bmbound3}), 
and therefore do not satisfy all microphysical constraints, unless one allows Vol($X_5$) (the dimensionless volume of the base of the warped throat geometry) to be very small.
In fact, we find that ${\rm Vol}(X_5) \lesssim 10^{-4}$ is required to prevent overshoot solutions. This is to be compared with the 
generic value of ${\rm Vol}(X_5) = \pi^3$ that we have used in our analysis. 
For parameters which do satisfy the constraints, the extension of the attractor 
solution does not occur; as proved in Appendix \ref{sec:slowroll}, the DBI attractor is close to the slow-roll limit
throughout once we impose the field range bound.

We find that known theoretical limits on the initial conditions phase space do not shrink it 
sufficiently to rule out all the 
overshoot trajectories. The difference from Ref.~\cite{Underwood:2008} is due to
Eq.~$(3.40)$ of that work missing a factor of $T_3^{-1}$, as can be seen on 
dimensional grounds. If the maximum initial momentum is $\Pi_\mathrm{max}$ then, with the revised bound given in Eq.~(\ref{eq:KKbound}), we have
\begin{equation}
		  \Pi_\mathrm{max} \propto D^{1/4}\,,
		  \label{eq:schematicmaxpi}
\end{equation}
where $D$ is a measure of the warped string scale at the tip of the throat (which determines the warped 3-brane tension at the tip).
On the other hand, the amplitude of the primordial power spectrum is proportional to $Ds$, as shown by 
Eq.~(\ref{eq:powerspec}). Here, $s$ parametrizes the ratio of energy density localized in the bulk relative to the energy density of the brane-antibrane system in the throat.
If the inflationary energy is dominated by the brane-antibrane tension 
[corresponding to $s\sim 1$ in Eq.~(\ref{eq:baumannpotential0})], the warped string scale 
at the tip constrains both the range of the initial momentum and the amplitude of the primordial 
power spectrum. Because the latter is tightly constrained by observations, we cannot solve the 
overshoot problem. 

If we allow the inflationary energy to be dominated by bulk energy
(corresponding to $s \gg 1$ in Eq.~(\ref{eq:baumannpotential0})), we break this correlation. 
The kinetic energy bound is still given by Eq.~(\ref{eq:schematicmaxpi}), and allows us to solve 
the overshoot problem provided $s \gtrsim 10^{20}$ (since the normalization of the power spectrum constrains the combination $D s$, this is equivalent to a very small value of $D$, the warped string scale at the tip). However, the bulk energy remains after the end 
of inflation. In this case, it may be challenging to explain how the Universe attains a state with small cosmological constant after inflation, {\it i.e.}~reheating would have to be addressed carefully.

The addition of an initially large, negative spatial curvature can mitigate overshoot \cite{Freivogel} by increasing the Hubble
friction. In Ref. \cite{Freivogel}, the inflaton gains kinetic energy by rolling down a steep potential after starting at
rest through a tunnelling event. In our formulation, we make no specific assumption about the mechanism by which initial conditions 
are set, only that they have to obey the microphysical bounds (\ref{eq:bmbound3}) and (\ref{eq:KKbound}), and the magnitude of the
kinetic term allowed by the backreaction bound is typically significantly larger than in the specific mechanism of \cite{Freivogel}. 
For large kinetic energy, the friction due to curvature is less effective. We found that the initial contribution of curvature to 
the Hubble parameter had to be larger than the contribution of the scalar field to have a significant effect. 
The largest initial kinetic energy for which the curvature term was effective was far below the microphysical bound; 
about $\Pi \sim 10^{-6}$. The equivalent figure without curvature is $\Pi \sim 10^{-7}$. This is not a significant increase in terms  
of resolving overshoot, and so we have neglected the effects of spatial curvature.

In order to determine which regions of parameter space suffered most from overshoot, we excluded the parts of the initial conditions space which always led to overshoot, and computed the proportion of the remaining space that led to successful inflation. We found parameter sets with a blue tilt $\ns > 1$ and a small warping scale suffered least from overshoot. In the inflection point case, avoiding overshoot to the same degree as the quadratic case required very blue spectra, inconsistent with observational limits.

Our analysis of the overshoot problem relied on currently available microscopic bounds on the inflaton field value $\phi_{\rm max}$ (from the geometric field range bound) and momentum $\Pi_{\rm max}$ (from backreaction considerations). We found that these bounds were too weak to allow a solution of the overshoot problem. However, it is conceivable that a future analysis could reveal stronger constraints on $\phi_{\rm max}$ and $\Pi_{\rm max}$. We therefore ask what bounds would be sufficient.
Reducing $\phi_{\rm max}$ while keeping $\Pi_{\rm max}$ fixed would make the overshoot problem more severe, since this provides less field space for dissipating excess kinetic energy.
The most promising avenue for solving the overshoot problem is therefore finding a stronger microscopic bound on the initial inflaton momentum. Quantitatively, our analysis showed that the new bound on $\Pi_{\rm max}$ would have to be at least 4 orders of magnitude stronger than the currently applied limit. Solving the overshoot problem for parameters giving a red tilt would require an improvement of 5 orders of magnitude. These conclusions are not sensitive to small perturbations of the potential shape away from the $\sim 60$ observationally-relevant $e$-folds of inflation.

As a by-product, our analysis recovered known observational predictions of quadratic brane inflation \cite{Bean:2007, Firouzjahi:2005},
and computed new predictions for the inflection point case. The model parameter space consistent with microphysical and observational restrictions predicts
unobservably small levels of both tensor modes and primordial non-Gaussianity, but a potentially observable cosmic string tension, in both the 
quadratic and inflection point cases.
We have also shown analytically that the DBI attractor solution will always slow-roll, and thus trajectories near it will generate nonobservable levels of non-Gaussianity. 

\acknowledgments We thank Bret Underwood and Lindsay King for useful discussions, Ben Freivogel for useful comments and 
Duncan Hanson for proofreading. 
DB thanks Igor Klebanov and Mikael Smedback for discussions and initial collaboration.
SB is supported by STFC. 
HVP is supported in part by Marie Curie Grant No. MIRG-CT-2007-203314 from the European Commission, and 
by STFC. 
The research of DB is supported by the David and Lucile Packard Foundation and the Alfred P.~Sloan Foundation, as well as the Center for the Fundamental  
Laws of Nature and the Institute for Theory and Computation at  
Harvard University.  
HVP thanks
the Galileo Galilei Institute for Theoretical Physics for their hospitality
and the INFN for partial support during the completion of this work.

\appendix

\section{Non-Relativistic DBI Attractor}\label{sec:slowroll}

For the DBI kinetic term and the potential used in this paper, significantly relativistic trajectories 
with $\gamma > 1.1$ are known to be ruled out by the field range bound  \cite{Baumann:2006} (see \cite{Bean:2007} for a numerical study). 
Here we prove analytically that the inflationary attractor is always close to the slow-roll limit, 
with $\gamma\approx 1$.

To describe the DBI inflationary attractor found by Ref.~\cite{Underwood:2008}, 
we define a new variable, 
\begin{equation}
  \chi \equiv -\dot{\phi}\sqrt{f} \,,
  \end{equation}
  such that
  \begin{equation}
  \dot{\chi} = \frac{\sqrt{f}}{\gamma^3}\left[V'-3H\gamma\chi-\frac{f'}{f^2}(\gamma-1)\right]\,.
  \label{eq:inflatattr}
\end{equation}
This has an attractor solution when $d\chi/dt \approx 0$:
\begin{eqnarray}
		  \chi_{\mathrm{Attr}} &=& \frac{(A+B)\sqrt{1+A^2+2AB}-B}{(1+(A+B)^2)} \nonumber\\
  &<& A +B\left(\frac{\sqrt{1+A^2+2AB}-1}{1+(A+B)^2}\right)\,,
  \label{eq:inflatattrsol}
\end{eqnarray}
where we have defined
\begin{eqnarray}
  A &\equiv& \frac{V'\sqrt{f}}{3H} \,,\label{equ:Aparam}\\
  B &\equiv& \frac{f'}{3H f^{3/2}}\,.
  \label{eq:attrparams}
\end{eqnarray}

If $A^2+2AB \ll 1$, then we have $\chi_{\mathrm{Attr}} \sim A$. The definition of $\gamma$ gives
\begin{equation}
  \gamma = (1-\chi^2)^{-1/2}\,.
  \label{eq:gammachi}
\end{equation}
To show that the attractor solution is nonrelativistic, it suffices to show that $A \ll 1$, 
irrespective of the size of $B$. To see this, we shall consider only the part of the attractor 
solution which depends on $B$, assuming that $A$ is small, and bracket the behavior in the two limiting cases. 
If $B$ is small enough that $A B$ is also small, then 
\begin{equation}
  \chi-A \ <\  \frac{AB^2}{1+(A+B)^2} \ < \ A \,.
\end{equation}
To derive the final inequality, note that this term is largest when $B \gg 1$.

If, on the other hand, $B > 1/A$, so that $A B$ is also large, we have
\begin{equation}
 \chi -A \ < \  \frac{\sqrt{2A}B^{3/2}}{1+(A+B)^2} \sim \sqrt{\frac{2A}{B}} \ < \ \sqrt{2} A\,.
\end{equation}
The second equality is because $A B \gg 1$, which implies that $A > B^{-1}$. In the worst case, 
therefore, the speed on the attractor is bounded by a small multiple of $A$.

We must now show that $A$ is small. Using (\ref{equ:Aparam}) we have the following upper bound
\begin{equation}
  A \ <\ \frac{\sqrt{f}V'}{3\sqrt{V}} \,.
  \label{eq:attrA}
\end{equation}

This can clearly be large if $f$ is allowed to grow without limit, but the field range bound prevents this.
Eq.~(\ref{eq:bmbound3}) can be recast in the form \cite{Peiris:2007} (in units where $M_{\rm pl} \equiv 1$)
\begin{equation}
		  \frac{\mathrm{Vol}(X_5)}{\pi} \int_{\phie}^{\phi_\mathrm{UV}} d\phi\ \, \phi^5 f(\phi)\ < \ 1 \,,
\end{equation}
which implies, for the AdS warp factor
\begin{equation}
  \frac{\pi^2}{2}\lambda\phi^2_\mathrm{UV} < 1 \,,\quad
  \Rightarrow \quad \sqrt{f} < \frac{\sqrt{2}}{\pi\phi_\mathrm{UV}\phi^2} \,.
\end{equation}

Substituting this and Eq.~(\ref{eq:baumannpotential0}), the form of the potential, into Eq.~(\ref{eq:attrA}) and 
dropping terms of order unity gives
\begin{equation}
  A \ <\  \frac{\sqrt{Ds}\left(2\phi +\frac{C D}{\phi^5}\right)}{\phi_\mathrm{UV}\phi^2\pi\sqrt{1+\frac{1}{3}\phi^2-a_\Delta \phi^{\Delta}-\frac{C D}{\phi^4}}}\, .
  \label{eq:attrisslowrolldemonstrated}
\end{equation}

This is much less than $1$ unless $\frac{D}{\phi^4} \sim 1$, by which time inflation has already ended. 
Thus the attractor is slow-roll, and as long as the inflaton moves on the attractor, we will have $\gamma \approx 1$, in 
agreement with the numerical results. 
\vfill

\bibliography{braneinflation}

\end{document}